%% file: main.tex
\documentclass[lettersize,journal]{IEEEtran}

\usepackage{mathptmx}      
\usepackage{amsmath,amssymb,amsfonts}
\usepackage{algorithmic}
\usepackage{algorithm}
\usepackage{graphicx}
\usepackage{svg}
\usepackage{nccmath}
\newcommand{\rset}{\mathbb{R}}
\newcommand{\argmax}{\arg\max}

\title{Quick unsupervised hyperspectral dimensionality reduction for earth observation: a comparison } 

\author{
Daniela Lupu$^{1}$, Joseph L. Garrett$^{2,3}$, Tor Arne Johansen$^{2,3}$, Milica Orlandic$^{2,4}$, Ion Necoara$^{1,5}$ 
	\thanks{*The research leading to these results has received funding from:  NO Grants 2014–2021, under RO-NO-2019-0184, project ELO-Hyp, contract no. 24/2020; UEFISCDI, Romania, under PNIII-P4-PCE-2021-0720, project L2O-MOC, no. 70/2022;  Research Council of Norway and industry partners under the Green Platform project, grant no. 328724.}
	\thanks{$^{1}$Automatic Control and System Engineering Department, National University of Science and Technology Politehnica  Bucharest, 060042 Bucharest, Romania, Emails: {\tt\small \{daniela.lupu, ion.necoara\}@upb.ro.}}%

	\thanks{$^{2}$ Center for Autonomous Marine Operations and Systems, Norwegian University of Science and Technology, Trondheim, Norway, Emails:  \{\tt\small joseph.garrett, tor.arne.johansen, milica.orlandic\}@ntnu.no.}
    \thanks{$^{3}$  Department of Engineering Cybernetics}
    \thanks{$^{4}$  Department of Electronic Systems}
     \thanks{$^5$ G. Mihoc-C. Iacob Institute of Mathematical Statistics and Applied Mathematics of the Romanian Academy, 050711 Bucharest, Romania.}
    }

\begin{document}

\maketitle
\input{Introduction}

\input{LinearMethods}

\input{Evaluation}
\input{Discussion}
\input{Conclusion}

\section{Acknowledgements}

We thank Sivert Bakken for proofreading a draft of the manuscript. The data are available from the NTNU smallsat lab, currently by request but the general-access HYPSO-1 data browser should soon be available. Code for the algorithms used here is available at https://github.com/ELO-Hyp/GUI-Hyperspectral-images .

\bibliographystyle{unsrt}
\bibliography{Bibliography}

\end{document}

%% file: Introduction.tex
\begin{abstract}
Dimensionality reduction can be applied to hyperspectral images so that the most useful data can be extracted and processed more quickly. 
This is critical in any situation in which data volume exceeds the capacity of the computational resources, particularly in the case of remote sensing platforms (e.g., drones, satellites), but also in the case of multi-year datasets.  Moreover, the computational strategies of unsupervised dimensionality reduction often provide the basis for more complicated supervised techniques. 
Seven unsupervised dimensionality reduction algorithms are tested on hyperspectral data from the HYPSO-1 earth observation satellite. Each particular algorithm is chosen to be representative of a broader collection.  
The experiments probe the computational complexity, reconstruction accuracy, signal clarity, sensitivity to artifacts, and effects on target detection and classification of the different algorithms. 
No algorithm consistently outperformed the others across all tests, but some general trends regarding the characteristics of the algorithms did emerge. 
With half a million pixels, computational time requirements of the methods varied by 5 orders of magnitude, and the reconstruction error varied by about 3 orders of magnitude. 
A relationship between mutual information and artifact susceptibility was suggested by the tests. 
The relative performance of the algorithms differed significantly between the target detection and classification tests.
Overall, these experiments both show the power of dimensionality reduction and give guidance regarding how to evaluate a technique prior to incorporating it into a processing pipeline.
 
\end{abstract}

\section{Introduction}

\begin{figure*}[tb]
    \centering
    \includegraphics[width=1.0\textwidth ]{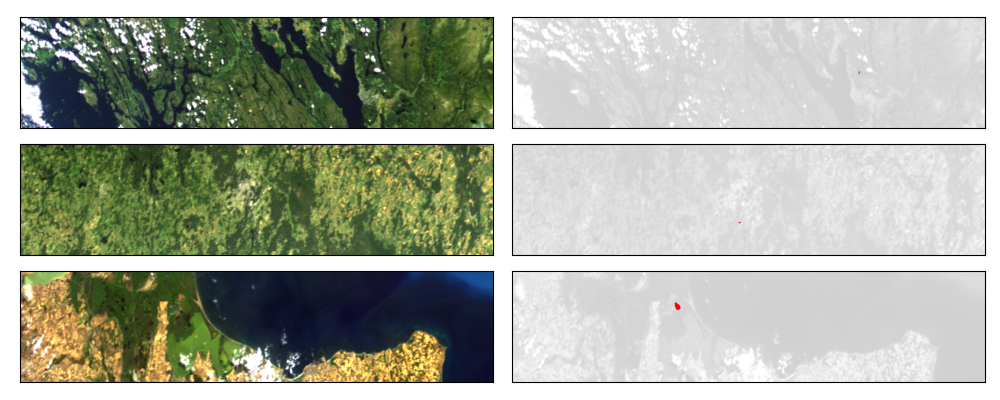}
    \caption{Three HYPSO-1 scenes used in the analysis: Trondheim, Norway August 23rd, 2022 (top), Vilnius, Lithuania July 22nd, 2022 (middle), Danube River Delta, Romania July 16, 2022 (bottom). The left side shows representations in red (651 nm), green (564 nm) and blue (458 nm), while the right side highlights the lakes which are used as targets in the target detection evaluation in red.}
    \label{fig:HYPSO_scenes}
\end{figure*}

Hyperspectral imagers are characterized by the high dimensionality of the data they collect, relative to multispectral cameras. 
While such data offer a more complete description of the world, their complexity can limit their utility.
Impediments to the analysis of high-dimensional data include the computational cost of reading them and a subtle decrease in accuracy as the dimensionality increases \cite{hughes1968mean}. 
Therefore, it is often useful to find a representation of the data with fewer dimensions, but which retains most aspects of the original data.
The transformation of high dimensional data $X$, with $N$ pixels that each have $d$ bands, into a representation with $r$ dimensions, where $r<d$, is known as dimensionality reduction. 

In the past two decades, a number of satellite missions have shown the value of hyperspectral remote sensing data, but their observations have suffered either from low temporal resolution or low spatial coverage \cite{folkman2001eo,barnsley2004proba,lucke2011hyperspectral,liu2019advanced,esposito2019orbit,cogliati2021prisma,jervis2021ghgsat,tanii2022orbit,bakken2023hypso,storch2023enmap}.
This has led some space agencies to plan larger missions which aim to regularly collect hyperspectral data from the entire earth \cite{buschkamp2023chime,dierssen2023synergies}. 
The storage and transmission limitations of these platforms, relative to the amount of data they collect has led to interest in assessing which algorithms are appropriate for such platforms \cite{alcolea2020inference}. 
Dimensionality reduction will be critical both to the retrieval and analysis of this data. 

A wide variety of tasks benefit from dimensionality reduction: target detection, classification, clustering, change detection, anomaly detection, data fusion, visualization, scene characterization (intrinsic dimensionality estimation), denoising, or compression \cite{harsanyi1994hyperspectral,chang2004estimation,farrell2005impact,ortiz2006change,du2007hyperspectral,li2011locality,fowler2011anomaly,liu2012unsupervised,johnson2012autogad,palsson2014model,falco2015spectral,feng2015random,luo2016minimum,pouyet2018innovative,danielsen2021self,chen2021component,cawse2023surface}.
However, recent comparisons of dimensionality reduction techniques tend to evaluate their performance on either target detection or classification \cite{chen2011effects,pu2016comparing,bakken2019effect,kumar2020feature,rasti2020feature,hong2021interpretable,li2022dimensionality}.
The analysis here expands the comparison in order to assess not only performance on a particular task, but to characterize the components which are found by each method and how they relate to each other. 
This broader focus will clarify which algorithms would be appropriate to incorporate into different parts of a processing pipeline. 
In particular, two of the criteria, time requirements and reconstruction accuracy, are particularly pertinent to platforms on which the computational resources are limited relative to the amount of data to be analyzed. 


In this work, we compare 7 methods of performing dimensionality reduction and subject them to a variety of different tests which are designed to evaluate their suitability for on-board processing systems. 
Each method is selected to represent the characteristics of a broader set of DR algorithms. 
Five of the methods perform a linear transformation of the data, which ensures linear computational times once the transformations are found. 
Two nearly-linear methods are included in order to identify any apparent limitations brought by linearity. 
Techniques which rely on spatial information have not been part of the analysis because they require the entire image, and thus cannot accommodate tests which train the DR from a random selection of pixels. 
Moreover, to use spatial analysis reliably in-flight would require accurate realtime hyperspectral georeferencing, which is a challenge in itself for pushbroom hyperspectral sensors. 
The tests are performed using data from scenes captured by  HYPSO-1 small satellite (Figs. \ref{fig:HYPSO_scenes} and \ref{fig:classification_scenes}). 


A description of the evaluated DR methods, together with a perspective of how they represent the wide range of hyperspectral DR is presented in Section \ref{sec:methods}. The motivation, description, and results of each of the experiments is presented in Section \ref{sec:experiments}. 
Some trends observed in the results are discussed in Section \ref{sec:discussion}.
Overall patterns and possible directions for future research are noted in section \ref{sec:conclusion}.





%% file: LinearMethods.tex

\section{Dimensionality reduction methods}
\label{sec:methods}

\noindent We consider a hyperspectral image represented as a matrix $X \in \rset^{N \times d}$, where $d$ is the number of bands and $N$ is the number of pixels. 
Linear DR methods map the observed data of dimension $d$ into a subspace of lower dimension $r \ll d$ using a linear transformation, i.e.,
\begin{equation}\label{eq:lm}
    \hat{X} = X W, 
\end{equation}
where $\hat{X} \in \rset^{N \times r}$ is the reduced data matrix and $W \in \rset^{d \times r}$ is the linear projection operator.
There are different approaches to finding the matrix $W$, each with a different technique of extracting  useful information.
In the following subsections, we present five linear DR methods.
Two additional non-linear methods, which generate transformations based on  more complicated functions than  \eqref{eq:lm} to map pixels to the reduced subspace, are presented as contrasts with the linear methods.   


\subsection{Principal Component Analysis (PCA)}
Principal Component Analysis (PCA) extracts information by finding the orthogonal linear transformation matrix $W$ that maximizes the variance of it's components \cite{hotelling1933analysis}, \cite{AbdWil:10}. 
The data are projected into the subspace spanned by the eigenvectors of the covariance matrix. 
To reduce the dimension of the projected data, one selects $r$ eigenvectors corresponding to the largest $r$ eigenvalues. 
The explicit computation of the covariance matrix is not necessary, as one can apply the singular value decomposition (SVD) directly to the observed data $X$ after centering (see Algorithm \ref{alg:PCA}). 

PCA represents a broad group of techniques that use singular value decomposition (SVD) to determine the components. 
PCA itself is the most popular of these techniques, as it maximizes the explained variance and in its simplest form has no tunable parameters beyond the number of retained bands \cite{jolliffe2016principal}. 
Maximum noise fraction combines SVD with an estimate of the noise \cite{gordon2000generalization,bioucas2008hyperspectral}.
Several approximate versions of PCA have also been developed, to increase its computational speed. 
The column-sampling version is incorporated in part of the analysis here to test whether it reduces the time required for processing hyperspectral data \cite{halko2011finding,HomMcD:16}. 
New developments such as tensor and sparse PCA maintain these techniques as an active area of research \cite{zou2006sparse,sun2019lateral,wang2023hyperspectral}. 
Sometimes it is also referred to as the Karhunen–Loève transform as well, particularly when it is incorporated into compression~\cite{hao2003reversible,penna2007transform}. 

PCA is a common tool for hyperspectral image analysis and can form the basis for real-time dimensionality reduction \cite{vitale2017fly,ortiz2019evaluating}.
It has also been combined with image segmentation, in the form of superPCA \cite{jiang2018superpca}.
However, PCA has received some scepticism regarding its effectiveness in hyperspectral image processing \cite{farrell2005impact,prasad2008limitations}. 

\noindent 

\begin{algorithm}[H]
	\caption{ PCA} \label{alg:PCA}
	\begin{algorithmic}
	\STATE 1. Center the data by removing the mean of each column: $X = X - \mathbb{E}[X]$
	\STATE 2. Construct matrix $Y = \frac{1}{\sqrt{N-1}} X$
	\STATE 3. [U, $\Sigma$, W] = svd(Y)    \hspace{0.1cm}
	\STATE 4. Select the first $r$ rows in $W$.
	\end{algorithmic}
	\label{alg_fastICA}
\end{algorithm}

Although SVD is more numerically stable when SVD is applied directly to the original data, one may prefer using the covariance matrix to reduce its memory requirements.
SVD has a computational complexity of order $\mathcal{O}(N^2d+d^3)$ for $N>d$ and approximately $\mathcal{O}(N^3)$ for $N \approx d$, thus yielding  memory and processing issues in the big data context.
In order to address these problems, strategies to \textit{approximate} PCA were developed in the literature \cite{ghashami2016frequent}, \cite{HomMcD:16}. In the column sampling approximations \cite{HomMcD:16}, the first step is to compute the covariance matrix:
\begin{equation}\label{eq:cov}
S =\frac{1}{N} X^{T}X = \begin{bmatrix} S_{11} \,\, S_{12}\\ S_{21} \,\, S_{22}\end{bmatrix}, 
\end{equation}
\noindent where the last equality represents a partition of $S$ such that  $S_{11} \in \mathbb{R}^{r \times r}$ and $S_{21} = S_{12}^T$. 
Note that $S_{11} = X_1^TX_1$, where $X_1$  contains $r$ columns of $X$. The $r$ columns are selected randomly.
Let us denote with $L(S) = \begin{bmatrix} S_{11}^T\,\, S_{21}^T\end{bmatrix}^T \in \mathbb{R}^{N \times r}$, and consider the spectral decomposition of $S_{11} = V(S_{11}) \Lambda(S_{11}) V(S_{11})^T$,
where $\Lambda(S_{11})$ is a diagonal matrix that contains the eigenvalues of $S_{11}$. 
Then, the low rank approximation of $S$ is 
\begin{align}
    S \approx \begin{bmatrix} S_{11}\\S_{21}\end{bmatrix} S_{11}^{\dagger}  \begin{bmatrix} S_{11}^T\,\, S_{21}^T \end{bmatrix},
\end{align} 
where $S_{11}^{\dagger}$ is the Moore-Penrose pseudo inverse of $S_{11}$. 
\noindent The eigenvalue decomposition of $L(S)$ leads to an alternative expression for $S$:
$$
S \approx U(L(S)) \Lambda(L(S)) U(L(S))^T,
$$
which yields the eigenvectors $W = U(L(S))$. The computational complexity for column sampling approximation is $\mathcal{O}(Nrd + d^2r)$, which is  $N$ order smaller than  classical PCA. 

\begin{figure}
    \centering
    \includegraphics[width=0.48\textwidth]{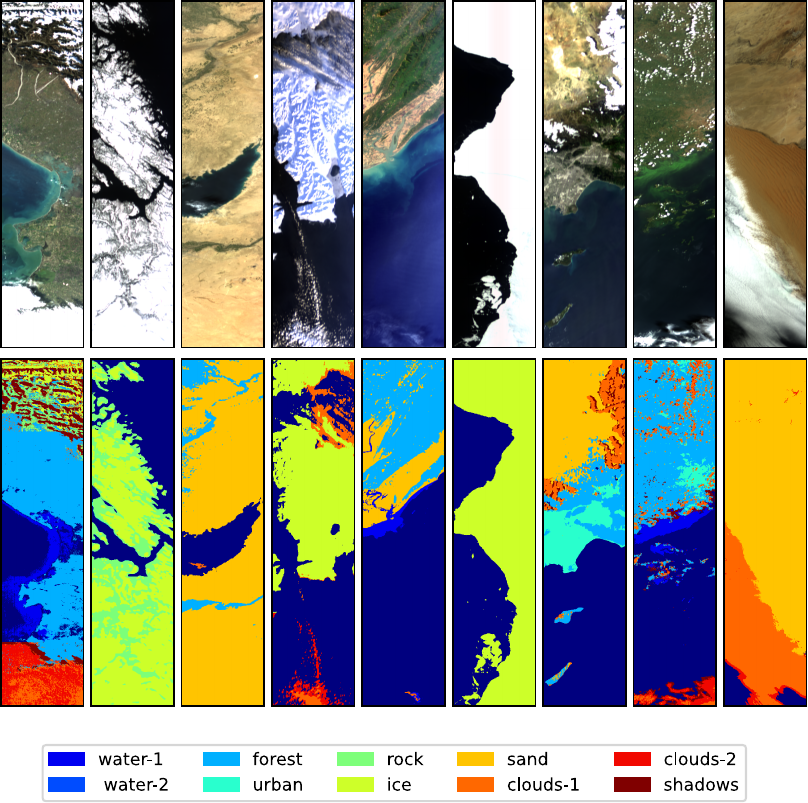}
    \caption{The additional HYPSO-1 scenes that are used for the classification test.}
    \label{fig:classification_scenes}
\end{figure}

\subsection{Independent Component Analysis (ICA)}

ICA represents a statistical dimensionality reduction technique.
The objective functions here include statistical properties of the normalized distribution, such as kurtosis or negentropy. 
FastICA is one of many ICA variants, which identifies components that follow a non-gaussian distribution~\cite{falco2014study}.
Other statistical dimensionality reduction techniques are e.g.,  Least Dependent Analysis and Principal Skewness Analysis, although ICA is the most common~\cite{stogbauer2004least, geng2014principal}.
After early successes, the application of ICA was criticized because its assumption regarding independent sources conflicts with the sum-to-one constraint of the linear mixing model, although this challenge is not insurmountable~\cite{nascimento2005does,wang2006independent, wang2006applications}. 
Independent components can be fused into morphological attribute profiles \cite{dalla2010classification} or extended to deep techniques \cite{li2022deep} and labelled data \cite{villa2011hyperspectral}.
Several optimization algorithms have recently been developed to increase the convergence speed of ICA~\cite{ablin2018faster,lupu2022stochastic,lupu2023convergence}.

ICA operates on the assumption that the reduced data (signals) is statistically independent \cite{HyvKar:01}.
Statistical indepedence is associated with non-gaussianity. 
By reverting the central limit theorem, which states that the sum of many signals will tend towards a gaussian distribution, one can suppose that a signal which is non-gaussian is the sum of very few independent signals.  There are many ways to quantify non-gaussianity in a signal, e.g.,  FastICA maximizes the negentropy.

\begin{equation} \label{eq:ica_pb}
\max _{\|w\|=1} \frac{1}{N} \sum_{i}^{N} g\left(x_i^T w\right),
\end{equation}
where $g$ is a non-quadratic function and $x_i^T$ represents the $i$th row of the matrix data  $X$. Three of the most used functions defining the negentropy are:
\begin{align*}
    &g(\hat{x}_i)=\frac{1}{\alpha} \ln \cosh (\alpha \hat{x}_i) \,\, \text{with} \,\,1 \leq \alpha \leq 2, \;\;\\ &  g(\hat{x}_i)=-e^{-\frac{\hat{x}_i^{2}}{2}}, \;\; \text{and} \;\;   g(\hat{x}_i)=\hat{x}_i^{3}. 
\end{align*}
To solve the nonconvex problem  \eqref{eq:ica_pb} one can employ an iterative optimization algorithms such as FastICA  \cite{hyv:99, HyvKar:01} or SHOICA \cite{lupu2022stochastic,lupu2023convergence}. Further, in Algorithm \ref{alg_fastICA} we describe the FastICA method for one unit, i.e., for finding one column in $W$:

\begin{algorithm}[H]
	\caption{ FastICA for one unit} \label{alg:fastICA}
	\begin{algorithmic}
	\STATE Choose a random vector $w_0$ and normalize it.
	\STATE \textbf{while}  $\delta \geq \epsilon $ :  
		\STATE \hspace{0.5cm } 1. Update rule: $$\tilde{w}_{k+1}=\mathbb{E}[x g^{\prime}\left( x^Tw_k\right)]-\mathbb{E}[ g^{\prime\prime}\left( x^T w_k\right)] w_k$$ \\
		\STATE \hspace{0.5cm } 2. Normalize: $w_{k+1}  \leftarrow \tilde{w}_{k+1}/\|\tilde{w}_{k+1}\|$\\
		\STATE \hspace{0.5cm } 3. Update stopping criterion $\delta = |w_{k+1}^T w_{k} - 1|  $\\
		\STATE \hspace{0.5cm } 4. $w_{k} \leftarrow  w_{k+1}$ and  increase $k$.
	\end{algorithmic}
	\label{alg_fastICA}
\end{algorithm}

\noindent  To avoid converging to the same $w$, after each iteration a decorrelation method is applied. Two primary methods used to decorrelate are the symmetric scheme in which a Gram-Schmidt-like decorrelation is applied to all the sources simultaneously \cite{HyvKar:01} and the deflation scheme based on sequentially estimating the signals one by one \cite{ZarCom:06}. The update rule of the FastICA algorithm is a Newton type iteration, where the expectancy can be approximated by the finite sum using empirical data.  In the derivation of the update rule, certain approximation were invoked, which necessitate a prerequisite preprocessing step, referred to as whitening.  Whitening entails applying a linear transformation $Q \in \rset_{d\times d}$ on $X$ such that $E[X^TX ] = I_d$. Several options for the matrix $Q$ exists in order to satisfy the aforementioned property.  However, in DR applications, a common preference is to choose it in a PCA manner:
$Q = \Lambda^{-\frac{1}{2}} C^{T}$, where $C \Lambda C^T  = X^TX$. FastICA may also identify local minima for  \eqref{eq:ica_pb}, so a viable alternative is SHOICA algorithm presented e.g., in \cite{lupu2022stochastic,lupu2023convergence}, which being a ascent method, will identify better points.  



\begin{table*}[h]
\centering
\caption{The tested dimensionality reduction techniques}
\begin{tabular}{|l|l|l|l|l|}
\hline
Name & Iterative & Data requirements  & Complexity  & Objective function  \\ \hline
PCA     & no           &  centered         &    $\mathcal{O}(N^2d+d^3)$ ($\mathcal{O}(Nr^2+r^3)$ )         &  maximize the explained variance                  \\ \hline
FastICA     & yes          &  whitened            &   $\mathcal{O}(rdN)$          &  maximize the negentropy                \\ \hline
OSP     & no           &                & $\mathcal{O}(Nd^2+d^3)$            &  no explicit objective function                    \\ \hline
LPP     & no           &                &   $\mathcal{O}(N^3)$          &  minimize distance between similar points \\ \hline 
VSRP    & no           &  centered         & $ \approx \mathcal{O}(rd)$            & preserve distances (on average) \\ \hline
NMF     & yes          &  positive entries &    $\approx \mathcal{O}(r^2(N+d) + rdN)$         &   minimize the distance to the original data \\ \hline
DBN     & yes          &  positive entries &  $\approx \mathcal{O}(2Ndr + Nd(r+d))$    &   minimize the distance to the original data                     \\ \hline
\end{tabular}
\end{table*}

\subsection{Orthogonal Subspace Projection (OSP)}

Orthogonal Subspace Projection (OSP) \cite{harsanyi1994hyperspectral}, represents a technique which is defined operationally rather than through an objective function.  OSP exists in both supervised and unsupervised forms and the latter, which is the automatic target generation procedure, is used here \cite{ren2003automatic}. 
In addition to being used as a DR technique on its own, it is also used to generate the initial conditions for other DR techniques \cite{wang2006independent,song2022bi}, or as  endmember identification and virtual dimensionality estimation \cite{chang2016comparative,chang2018review}. 
The recently-developed orthogonal projection endmember algorithm (OPE) follows the same structure as OSP, with some simplified steps to increase its computational speed \cite{tao2022fast}. From the OSP perspective \cite{chang:05} one pixel $x \in \mathbb{R}^{d}$ can be modeled as:
\begin{equation}\label{eq:osp_signal_model} 
x = \alpha t + \Omega \gamma  + v,
\end{equation}
where $t \in \mathbb{R}^{d}$ is the desired target spectral signature, $\Omega$ is the undesired target spectral signature matrix, $\alpha \in \mathbb{R}$ and $\gamma \in \mathbb{R}^{d-1}$ are abundance coefficients, and $v$ is white noise with zero mean and $\sigma^2$ variance. 
Separating the spectral signatures in this manner allows us to design an orthogonal subspace projector to remove $\Omega$ from the image pixel vector $x$. 
One such projector is:
$$
P = I - \Omega \Omega^{\dagger},
$$
where $\Omega^{\dagger} = (\Omega \Omega^T)^{-1}\Omega^{T}$. 
Applying this projection to \eqref{eq:osp_signal_model}, we nullify the undesired targets and suppress the noise:
 \begin{equation}\label{eq:osp_model}
      Px = Pt \alpha + Pv.
 \end{equation}
\noindent Upon the elimination of non-desirable targets, we employ a linear matched filter to track the intended targets, defined as $M_{t} x= t^Tx$. Thus, the OSP's $W$-matrix has the following expression:
$W = M_{t}P = t^TP$. 

\noindent Despite the apparent simplicity, finding the pseudo-inverse of $\Omega$ is computationally heavy.  Although this step involves calling the SVD routine (see Section 2.1), the complexity is relatively small: $\mathcal{O}(d^3)$.

\begin{algorithm}[H]
	\caption{ Orthogonal Subspace Projection (OSP) } \label{alg:osp}
	\begin{algorithmic}
	\STATE 1. Initialize $\Omega = [ \ ]$. 
        \STATE 2.  \textbf{for} i= 1: r \textbf{do}:
        \hspace{0.5cm}\STATE 2.1  Compute projector $P = I - \Omega \Omega^{\dagger}$ and apply it on $X$. 
        \hspace{1cm}\STATE 2.2 Compute:
        $ t_{i} = \argmax_{x_j, \,j=1:N} (Px_j)^T(Px_j)
        $
        \hspace{0.5cm}\STATE 2.3 Update $\Omega = [t_1, \cdots t_{i}]$\\
        \STATE 3. $W = \Omega$
	\end{algorithmic}
	\label{alg_lpp}
\end{algorithm}


\subsection{Locality Preserving Projection (LPP)}
Locality Preserving Projection (LPP)  represents a technique which relies on an adjacency graph to describe the relationships between pixels~\cite{HeNiy:03}. It is often grouped together with an assortment of non-linear DR techniques that similarly rely on the adjacency graph~\cite{yan2006graph,rasti2020feature}.
There are several ways to determine the adjacency graph without supervision, including $k$-nearest neighbors or separation thresholding, and several new unsupervised techniques have emerged by incorporation  regularization \cite{HeNiy:03,jiang2022unsupervised,chen2023unsupervised}. 
While LPP itself is unsupervised, it provides the basis for many supervised or semi-supervised techniques \cite{li2011locality,lunga2013manifold,zhai2016modified,li2018discriminant,rasti2020feature}. This method builds an adjacency graph $G$ incorporating neighborhood information of the data set. 
The implementation here relies on a $k$-nearest neighbors algorithm for defining adjacency. 
First the $k$ nearest (spectral) neighbors of each pixel are found. 
The adjacency graph, $G \in \mathbb{R}^{N\times N}$, is constructed so that $G_{ij}=1$ if pixel $i$ is one of the $k$ nearest neighbors of pixel $j$, and $0$ otherwise. 
Then, the Laplacian matrix of the graph is calculated as $L = D - G$, where $D$ is a diagonal matrix whose entries are column (or row, since $H$ is symmetric) sums of $H$, i.e., $D_{ii} = \sum_j G_{ji}$. Finally,  the projection matrix $W$ is found by solving the general eigenvector problem \eqref{eq:gep}. 
\begin{algorithm}[H]
	\caption{ Locally preserving projection (LPP) } \label{alg:lpp}
	\begin{algorithmic}
	\STATE 1. Construct the adjacency graph $G$ 
 
        \STATE 2. Compute the eigenvectors $w$ and eigenvalues for the generalized eigenvector problem: 
        \begin{equation}\label{eq:gep}
            X^TLX w = \lambda X^TDX w.
        \end{equation}
	\end{algorithmic}
	\label{alg_lpp}
\end{algorithm}
\noindent This linear transformation preserves local neighborhood information \cite{HeNiy:03}. Note that from a memory point of view the weight matrix can be very large. Moreover, step $2$ is the most expensive step, having a complexity of order $\mathcal{O}(N^3)$. To ameliorate this disadvantage, we use a random subset of the data to work with a smaller matrix. Thus, choosing $N_s \ll N$, the number of pixels, the complexity improves to $\mathcal{O}(N_s^3)$.  
In our numerical tests, we take $k=5$ nearest neighbors. 


\subsection{ Very Sparse Random Projection (VSRP)}
The random projection approach is a straightforward and computationally efficient strategy for reducing the dimensionality of data by not considering the structure of the data  \cite{LiHas:06}. 
According to the Johnson-Lindenstrauss lemma, if $W$ has independent and identically distributed entries,  then the inter-point distances within a high-dimensional space are approximately maintained in a lower-dimensional space, up to some level of distortion $\epsilon$.
In \cite{achlioptas2003database}, Achlioptas proposed using $W$ of the following form:
\begin{align}\label{eq:vsrp}
w_{i, j}=\sqrt{c} \begin{cases}1 & \text { with prob. } \frac{1}{2 c} \\ 0 & \text { with prob. } 1-\frac{1}{c} \\ -1 & \text { with prob. } \frac{1}{2 c},\end{cases}
\end{align}
where typically $c =3$. With this value for $c$, one can achieve a threefold speedup because only $\frac{1}{3}$ of the data needs to be processed (hence the name \textit{sparse random projection}). In \cite{LiHas:06}, the authors propose setting $c =\sqrt{d}$, for a very sparse random projection. VSRP \cite{achlioptas2003database,LiHas:06,martin2016hyperspectral,feng2015random} represents a DR method related to random projections, which can be used as a form of compressive sensing \cite{fowler2009compressive,du2011random}. 
These methods can reduce the dimensionality before the whole dataset is acquired because the projections are random and thus do not depend on the data themselves, unlike the other methods presented here. 
The papers which address random projections as a dimensionality reduction technique for hyperspectral data typically present it as a form of preprocessing for particular tasks such as target detection, (de)compression, classification, or unmixing \cite{feng2015random,fowler2009compressive,du2011random,fowler2011anomaly,menon2016fast,fox2016random,menon2017random}.

\begin{algorithm}[H]
	\caption{ Very Sparse Random Projection (VSRP) } \label{alg:vsrp}
	\begin{algorithmic}
	\STATE 1. Construct $W$ using \eqref{eq:vsrp} for $c = \sqrt{d}$
        \STATE 2. Reduced image $\hat{X} = X W$
	\end{algorithmic}
	\label{alg_lpp}
\end{algorithm}

\noindent This approach is very simple, having the complexity of order $\mathcal{O}(rd)$. 
However, $r$ needs to be relatively large for the level of distortion $\epsilon$ to be small. 



\subsection{Nonnegative Matrix Factorization (NMF)}

NMF factors a matrix $X$ into the product of two nonnegative matrices $W$ and $H$, so that $X\approx WH$ \cite{lee1999learning,gillis2014and}.
It is commonly used as a hyperspectral unmixing technique, rather than a DR technique \cite{FenHen:2022}. 
Unmixing differs from DR in that it imposes a sum-to-one constraint on each pixel so that the decomposition can be interpreted according to the linear mixing model. Without that constraint, the components found by NMF can be normalized so that the $L2$ norm of the columns of $W$ are set to one.
Then, the scaling problem, which plagues unmixing, vanishes.
NMF in this form has no need for regularization.  
Although the NMF decomposition is linear in each of its constituent matrices, it is considered a non-linear DR technique within the scope of this comparison because the function that maps a pixel to the reduced dimensionality generally requires non-negative least squares. 

There are many iterative algorithms in the literature to solve this nonconvex optimization problem, see e.g., \cite{chi2019nonconvex,ahookhosh2021multi,FenHen:2022,ChoLup:22}.
One efficient algorithm for solving this problem  is based on the  following multiplicative update rule \cite{lee2000algorithms}:
\begin{align}
H_{ij} &\leftarrow H_{ij} \frac{(W^{T}X)_{ij}}{(W^{T}WH)_{ij}} \\
W_{ki} &\leftarrow W_{ki} \frac{(XH^{T})_{ki}}{(WHH^{T})_{ki}},
\end{align}
where the indices denote each element. In the implementation of NMF used here, the initial $W$ and $H$ matrices are calculated using OSP (discussed above). The multiplicative update rule is then applied to the matrices $W$ and $H$ repeatedly until they reach a predefined convergence tolerance. Matrix 
$W$ is normalized after each update step.

\begin{figure*}[tb]
    \centering
    \def\svgscale{1.0}
    \includegraphics[width=1.0\textwidth ]{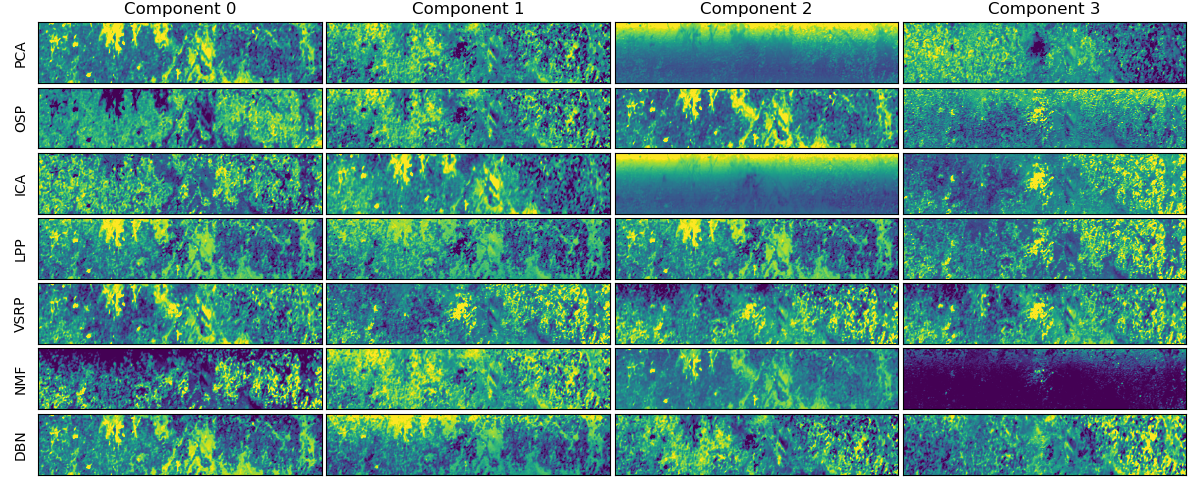}
    \caption{The Vilnius scene decomposed by PCA, ICA, OSP, LPP, VSRP, NMF and DBN}
    \label{fig:Trondheim_decompositions}
\end{figure*}

\subsection{Deep Belief Network (DBN)}

DBNs are a simple type of artificial neural network that are composed of a series of Restricted Boltzmann Machines (RBMs) \cite{hinton2006fast,chen2015spectral,ma2018unsupervised}. 
These non-linear networks represent a small part of the variety of neural networks which have been applied to hyperspectral data. 
DBNs are fast networks with some implementations designed for remote platforms \cite{boyle2023high}. 
The architecture used here has only one hidden layer. The outputs of the hidden layer are used as the reduced components. Here, the DBNs are trained as an autoencoder, in which one layer is the encoder:
\begin{align}
F(X) = S(W_{0} X + b_{0}), 
\end{align}
and another is the decoder:
\begin{align}
G(x) = S(W_{1} X + b_{1}),
\end{align}
so that $H(X) = G \circ F(X) \approx X$ after training, where we consider the activation function sigmoid $S(x) = (1+e^{-x})^{-1}$.
To facilitate quick training, the DBNs used in this paper are composed of one input layer, one hidden layer, and one output layer, with sizes $[B, D, B]$, where $B$ is the original number of bands in an image and $D$ is the reduced number of components. The training process is divided into two portions, a contrastive divergence pre-training and fine-tuning by gradient descent.  Contrastive divergence (CD) pretrains the network layer-by-layer as separate RBMs.  $F$ is then the map from a visible layer, $X$, to a hidden layer, $Y$.
Another function, $\bar{F}(Y) = S(W_{0}^{T} Y + c_{0})$ is defined to map the hidden layer back to the visible layer. 
Hence, $\bar{F} \circ F$ is  a RBM. 
CD then adjusts $W_{0}, b_{0},$ and $c_{0}$ iteratively:
\begin{align}
\Delta W_{0} &= \overline{X_{0} \otimes Y_{0} - X_{i} \otimes Y_{i}},\\
\Delta b_{0} &= \overline{Y_{0} - Y_{i}},\\
\Delta c_{0} &= \overline{X_{0}- X_{i}},
\end{align}
where $X_{i+1} = \bar{F}\big(F(X_{i})\big)$, $Y_{i+1} =F\big(\bar{F}(Y_{i})\big)$, $i$ denotes the length of the Markov chain used, and $\overline{X}$ denotes an average over several pixels. 
The networks are trained with momentum to reduce noise in the update process, and an L1 weight penalty is also included. 
After $F$ is trained, the same CD process is used to train $G$, using $Y_{0}$ as input. 
Finally, gradient descent is used to fine-tune the parameters. 
The pretraining proceeds for a number of iterations, while the gradient descent training repeats until a certain tolerance is reached. 
The weights of the network are initialized randomly. The computational complexity of CD per layer is $\mathcal{O}(N(rd+r+d)i)$ per epoch, where $i$ is the length of the Markov chain, which is linear in $N$.
The computational complexity of gradient descent per layer is similarly $\mathcal{O}(Nd(r+1))$, and thus also linear in $N$.
Here, $i=2$ for the tests.




%% file: Evaluation.tex
\section{Evaluation of DR methods}
\label{sec:experiments}

The dimensionality reduction algorithms are evaluated in 6 different tests:
\begin{itemize}
    \item Computation time 
    \item Reconstruction accuracy
    \item Independence of components
    \item Sensitivity to artifacts
    \item Target detection performance
    \item Classification performance
\end{itemize}

The first four tests characterize the internal properties of the algorithm without reference to any ground truth, while the final two tests characterize how data processed by the DR algorithms performs within subsequent analysis, when a ground truth is available. 
In order to emulate a remote platform which processes a single image at a time, the first tests investigate how the DR algorithms function on single images, and so three radiance images are used (see Figure \ref{fig:HYPSO_scenes}).
The final test considers the situation in which DR is applied to a larger dataset of 9 additional images, together with the original Trondheim scene, which are labelled and standardized by converting to top-of-atmosphere reflectance \cite{Roysland2023}. 
The tests are run on an Intel Core i7-4790S CPU (3.2GHz) with 16 GB RAM. For hyperspectral data analysis algorithms which are intended to be run in-flight,  computational speed is a critical property. 
Therefore, the first test evaluates the speed of the different algorithms. 
In general, the dimensionality process can be split into two parts: first the reduction is computed and, second, it is applied to each pixel in the image. 
Here the focus is on the first half of the process because the second portion of the process requires the same amount of time for all the linear techniques. 
A particular focus is placed on how time required to compute the different transformations scales with the number of pixels and number of output bands. 

One of the tasks that DR algorithms are commonly used for is lossy compression. 
Only some of the bands in the reduced dimensional space are retained to facilitate faster data transfer. Therefore, this paper evaluates how much variance is retained with bands from the different DR methods. In addition, we explore how much sub-sampling during the calculation of the transformation affects the amount to information which is retained in the bands. 

DR is often used to reduce the redundancy within hyperspectral data. Here, the redundancy between pairs of bands is characterized by the mutual information between them. 

Hyperspectral images are very susceptible to a number of artifacts, including line noise, smile and keystone, and wavelength-dependent blur. Therefore, we attempt to characterize how much the various DR techniques are affected by imaging artifacts, chiefly striping and smile. 

The most common applications of hyperspectral data are for target detection \cite{manolakis2013detection} or classification \cite{chen2014deep}. DR is often seen as a way to prepare data for these applications. By reducing the dimensionality and decorrelating the bands, it can improve the performance of target detection and classification algorithms.

\begin{figure}[tb]
    \centering
    \includegraphics[width=0.48\textwidth]{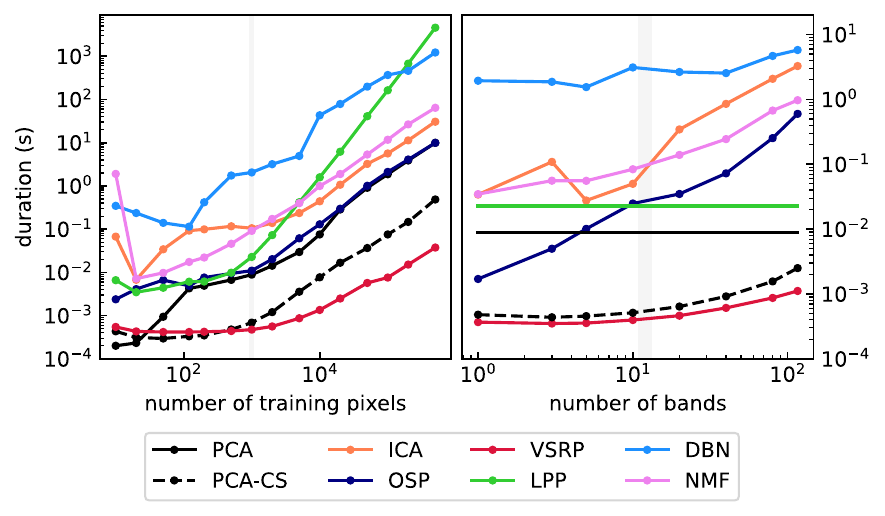}
    \caption{The median time required to compute the transformation with different numbers of pixels (left, 12 bands). The time to compute the transformation with a limited number of bands (right, 1000 pixels).} 
    \label{fig:computing_times}
\end{figure}

\subsection{Computation time}

Because time is among the strictest constraints in most computational settings, it is evaluated first. 
The time to calculate a particular transformation is recorded for a varying number of input pixels. 
Not only can the relative speed of the different algorithms be determined, but it is also possible to assess how the time scales with the number of training pixels, which relates to the computational complexity of the algorithm.
The time necessary to apply the linear transformations to the data does not depend significantly on the technique, about 12 $\mu$s per pixel for techniques which center the data, or about 10 $\mu$s for techniques which do not remove the mean.
Of the nonlinear techniques, DBN requires about 10 $\mu$s per pixel, but NMF requires nearly 100 $\mu$s per pixel.
The sparse vectors of VSRP are converted to dense vectors here, as the scipy implementation of sparse vectors requires about 3$\times$ longer to compute \cite{virtanen2020scipy}. 
However, in principal, a sparse implementation of VSRP should be faster than the dense implementation, so some acceleration is possible.  
Note that the column sampling variant of PCA discussed above is also tested in this experiment \cite{HomMcD:16}.

Some of the algorithms -- OSP, ICA, DBN, and NMF -- allow a subset of the bands to be calculated, while LPP and PCA compute all bands innately. 
For these techniques, the first speed test is performed with 12 bands, and the a second test is performed, in which a varying number of bands is calculated. 
Figure \ref{fig:computing_times} shows the results of these tests for the Vilnius scene. 
The training pixels are selected randomly, but are re-used for each method.
The results for the other scenes are similar.

\begin{figure}[tb]
    \centering
    \includegraphics[width=0.49\textwidth]{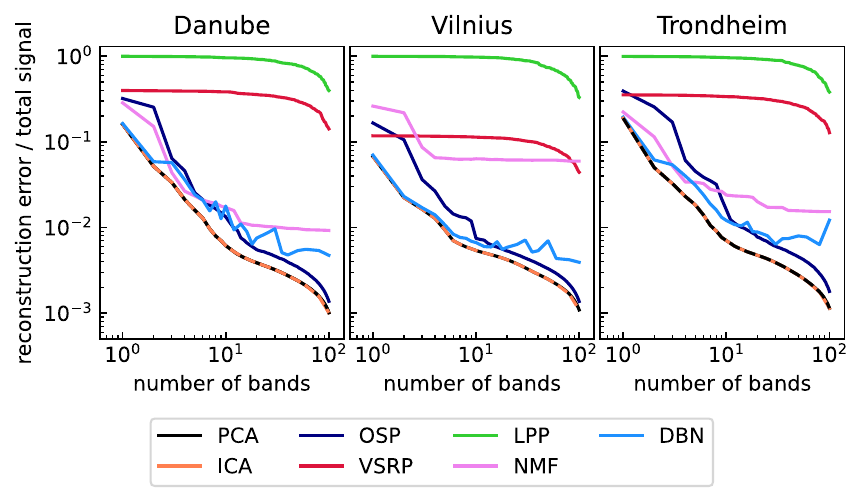}
    \caption{The reconstruction error of the three scenes with the different transformations. Note that because FastICA is a transformation of the PCA subspace, lies directly on the PCA curve. }
    \label{fig:FullPCAwwoclouds}
\end{figure}

\subsection{Reconstruction accuracy}

The goal of the reconstruction tests is to assess how accurately an entire scene can be reconstructed from the dimensionally reduced bands of the different techniques. 
The metric used is pixel-wise error in the reconstruction relative to the original pixel intensity:
\begin{equation}
E_{R} = \medop\sum_{i=0}^{NM}{ \frac{(X_{i} - F^{\dagger}F(X_{i}))^{2}}{X_{i}^{2}}},
\label{eq:rec_error}
\end{equation}
where $F$ is the transformation and $F^{\dagger}$ is used to reconstruct the data. 
$F^{\dagger}$ is the pseudoinverse of the matrix for the linear techniques, while NMF and DBN implicitly have reconstruction functions. 
The metric is 0 for perfect reconstruction, and approaches one when $F^{\dagger}\circ F$ vanishes. 
This metric is normalized pixel-by-pixel so that the spectral signatures of relatively dark pixels, such as water pixels, also contribute.

The reconstruction accuracy is first evaluated as a function of the number of bands the image is reduced to, on the three different scenes (Fig. \ref{fig:FullPCAwwoclouds}). 
Note that FastICA is a rotation of the PCA bands, and so they contain the same information when reconstructed, by design. 
The transformations PCA, ICA, VSRP each remove the mean from the data before processing and return it during the reconstruction, while the others do not, so the former start with a reduced $E_{R}$, even with 0 bands. 
Moreover, the linear methods, each being complete basis of the HSI image achieve $E_{R}=0$ when all bands are used, whereas the non-linear methods do not. 

\begin{figure}[thb]
    \centering
    \def\svgscale{1.0}
    \includegraphics[width=0.48\textwidth]{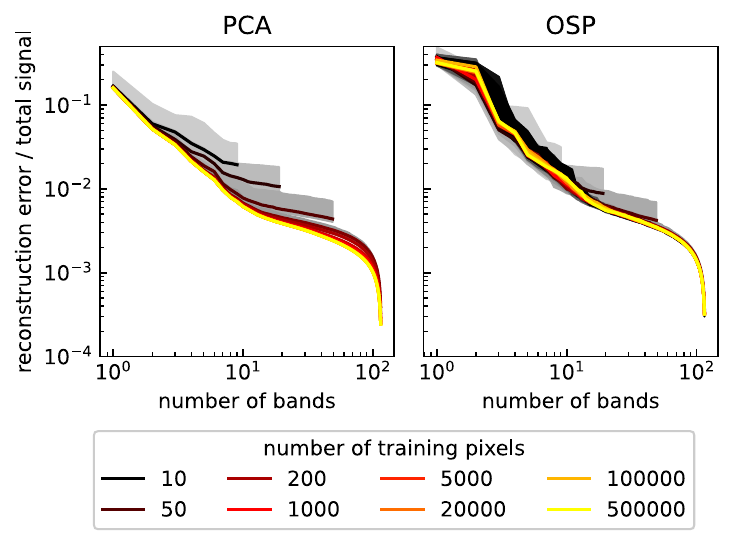}
    \caption{The reconstruction error using different numbers of pixels, for the PCA and OSP techniques. The solid line shows the median of 10 sets of pixels, and the shaded regions show the range.}
    \label{fig:FullPCA_subsamp_meanerr}
\end{figure}

There is the possibility that training with only a portion of the image, can also reduce the reconstruction accuracy of the components which are found, so that it accelerates computation. 
Therefore, the effect of the number of pixels on reconstruction accuracy is evaluated for the different methods. 
The results for the methods with the best accuracy in the first evaluation, PCA and OSP, are shown in Fig. \ref{fig:FullPCA_subsamp_meanerr}.

\subsection{Component independence}

The spectra recorded in hyperspectral imagery originate from many diverse sources. 
One goal of dimensionality reduction is to separate the signals originating from different sources. 
Here this is characterised through the mutual information (MI) of the different components found by each algorithm in each image. MI describes how much information in one random variable is present in another \cite{kraskov2004estimating}. 
Some of the DR techniques presented here, PCA for example, are constructed to eliminate covariance between the different components they produce. 
However, that does not imply that the components are independent. 
MI is used to detect more complex relationships between the components. 
MI is theoretically expressed in terms of the (joint) probability distributions of the different components. 
However, in this interpretation, the components are quantized samples drawn from these unknown distributions, which make a direct application of the distributional framework  sensitive to hyperparameters like bin size. 

\begin{figure}[tb]
    \centering
    \includegraphics[width=0.49\textwidth]{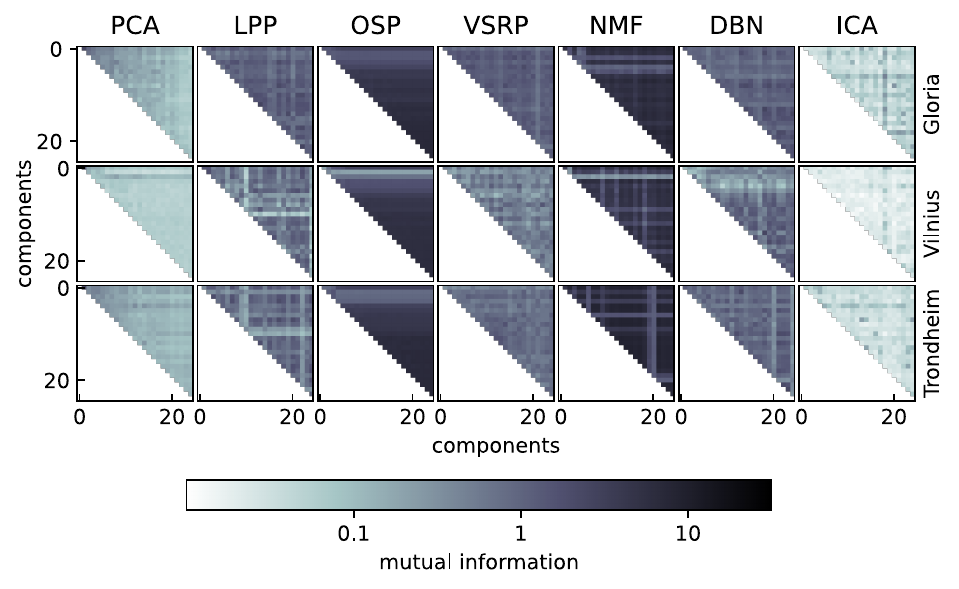}
    \caption{The mutual information of the images, computed between the bands of each DR technique.}
    \label{fig:reduced_MI}
\end{figure}

The MI is estimated through the $k$-Nearest Neighbors algorithm, which was developed to minimize the sensitivity on hyperparameters \cite{kraskov2004estimating,gao2018demystifying}.
Then, the mutual correlation between each pair of bands is computed for each scene, for a random sample of 43120 points (the same points for each algorithm).
The results are plotted as a heat map in Fig. \ref{fig:reduced_MI}, in which darker areas indicate greater mutual information.
For each type of transformation, the number of components is reduced to 25.
The number 25 was chosen to be large enough for trends between the components to become apparent, but small enough so that the mutual information did not become dominated by noise. 
For the techniques which produce ordered components (PCA, LPP, OSP), they are plotted in their natural order, otherwise the order is the arbitrary order in which the components are found. 
Because the mutual information is symmeteric with regards to the order of the components, only the upper-triangular part of the matrices are plotted.


\subsection{Sensitivity to artifacts}

Hyperspectral imagers are afflicted by a number of artifacts: smile, keystone, stray light, wavelength-dependent blur, among many others \cite{skauli2017feasibility}. 
These artifacts are systematic errors which differ from many types of noise in that they cannot be corrected by collecting more data or averaging pixels.  
Attempts to directly correct these artifacts after image acquisition require longer processing time and can introduce additional sources of uncertainty, while correcting them optically typically requires large and expensive optics.  
Therefore, a dimensionality reduction technique that is sufficiently robust against artifacts could shield downstream processing from their effects.

\begin{figure}[tb]
    \centering
    \includegraphics[width=0.49\textwidth]{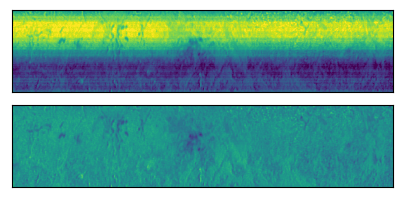}
    \caption{The 6th of 25 components from ICA on the Vilnius image before (top) and after (bottom) the along-track means have been removed, on the same colorscale. The ratio of across-track variance to total variance is 0.92.}
    \label{fig:stripes}
\end{figure}

A number of artifacts produce a distinct signature in the along-track direction of a push-broom imager. 
The test here refers to two particular artifacts: striping and smile.
Smile refers to a parabolic distortion of wavelength as a function of a pixel's spatial position across-track. 
Striping is a variation in the radiometric response as a function of the pixels spatial position. 
Both of these artifacts can manifest as lines in the along-track direction, as the error accumulates through each sequential spectrogram. 
An example of both artifacts in one component can be seen in Figure \ref{fig:stripes}.
The thin lines are examples of striping, while the stark contrast between the upper- and lower-halves of the figure is likely due to smile. 
Both of these artifacts are relevant to any camera that uses a push-broom image acquisition because they affect each sequentially acquired spectrogram in a similar way. 

Because these two sources of error manifest as along-track lines, we characterize them by calculating the proportion of the variance of each component which is retained in along-track lines. 
Then, the mean of each along-track line is removed (Figure \ref{fig:stripes}, bottom part), and the levelled variance, $V_{L}$,  is calculated. 
The across-track proportion of variance (ATPV) is then given by:
\begin{equation}
P_{A} = 1 - \frac{V_{L}}{V_{T}}.
\end{equation}
In the limit that a component is composed entirely of stripes, this metric will be one. 
However, even images without any stripes are expected to contain a non-zero ATPV. 
Therefore, the ATPV for each component is compared against the average of all bands for each scene. 
Note that the striping noise is present in the original data, not injected during the testing process, and thus is present in all the tests, including the reconstruction tests described above. While assessments of stripe noise rely on accuracy metrics, such as the peak signal-to-noise ratio for simulated data, it is more difficult to assess the striping in real data. 
Although it is common to rely on visual inspection, e.g. \cite{pande2011striping}, two metrics are occasionally used to assess striping in real data: the inverse coefficient of variation and the noise reduction coefficient \cite{chang2015anisotropic}. 
The former uses the standard deviation over a uniform area to estimate the stripe noise, while the latter compares the Fourier-transformed image before and after destriping at the frequencies which could correspond to striping. 
Unlike the inverse coefficient of variation, ATPV does not require a uniform region to be selected and unlike the noise reduction coefficient, it can directly evaluate a single component, rather than require a striped and de-striped pair. 
The authors of \cite{rogass2014reduction} define the Absolute Average of Highpass of Differences metric in order to investigate stripes more directly. 
The ATPV is similar to the Absolute Average of Highpass of Differences, except that the former does not require hyperparameters to be selected. 

\begin{figure*}[tb]
    \centering
    \includegraphics[width=0.98\textwidth]{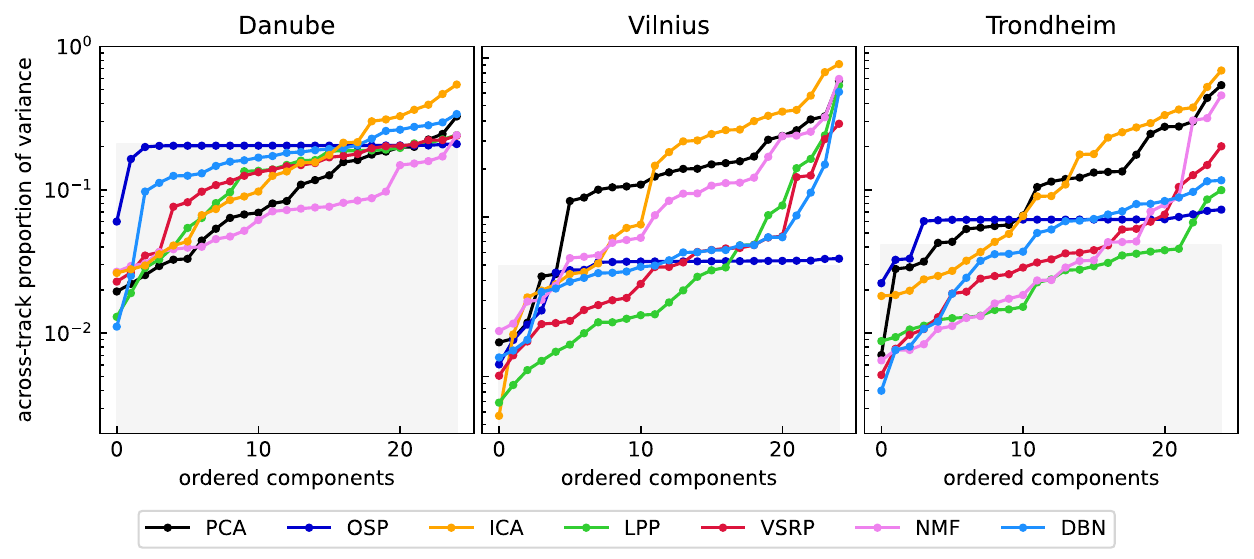}
    \caption{The across track proportion of variance, which characterizes the along-track striping artifact, for components calculated using the different DR technqiues.}
    \label{fig:alongtrack_variance}
\end{figure*}

The across-track proportions of variance for each method, for the same 25 components used in the above mutual information analysis, are plotted in Figure \ref{fig:alongtrack_variance}. 
Within each method, the components are reordered according to this proportion, for clarity. 
Note that there is a strong scene-dependence on these different sources of noise, because some features, such as the coast in the Danube Delta scene, are also characterized by systematic across-track variation. 
Although this is not a complete characterization of all noise sources, it does identify a number of components which are particularly affected by these two artifacts. 

\begin{table}[htb]
    \centering
    \begin{tabular}{c|c|c|r}
         Name of Lake & Location & Date & Pixels \\ \hline
         Dranov (Romania) & 44.9$^{\circ}$ N 29.1$^{\circ}$ E & July 16, 2022 & 362 \\
        Didžiulis (Lithuania) & 54.7$^{\circ}$ N 25.0$^{\circ}$ E & July 22, 2022  & 28\\
         Jonsvatnet (Norway) & 63.4$^{\circ}$ N  10.6$^{\circ}$ E & August 23, 2022 & 28\\
    \end{tabular}

    \caption{Lakes used for target detection}
    \label{tab:HICOscenes}
\end{table}

\subsection{Target Detection}

Because the mission of HYPSO-1 is oriented towards water monitoring, one lake in each image is selected as a target. 
The in-scene pixels which correspond to each lake are averaged in order to determine a target spectra. 
Two different target detection algorithms are used, the spectral angle mapper (SAM) and the Adaptive Cosine Estimator (ACE) \cite{manolakis2013detection}. 
The spectral angle mapper is defined as:
\begin{equation}
    \text{SAM}(x) = \frac{x^{T} x_{t}}{(x^{T} x)^{1/2}(x_{t}^{T} x_{t})^{1/2}},
\end{equation}
where $x_{t}$ is the target spectrum. 
The spectral angle mapper is similarly defined as:
\begin{equation}
    \text{ACE}(x) = \frac{x^{T} C^{-1} x_{t}}{(x^{T}C^{-1}x)^{1/2}(x_{t}^{T}C^{-1}x_{t})^{1/2}},
\end{equation}
where $C$ is the covariance matrix over the scene. 
For both detectors, a value closer to one is a more likely detection. 
Typically, some threshold is set so that values larger than the threshold are considered to be likely targets.

\begin{figure*}[tb]
    \centering
    \includegraphics[width=0.6\textwidth]{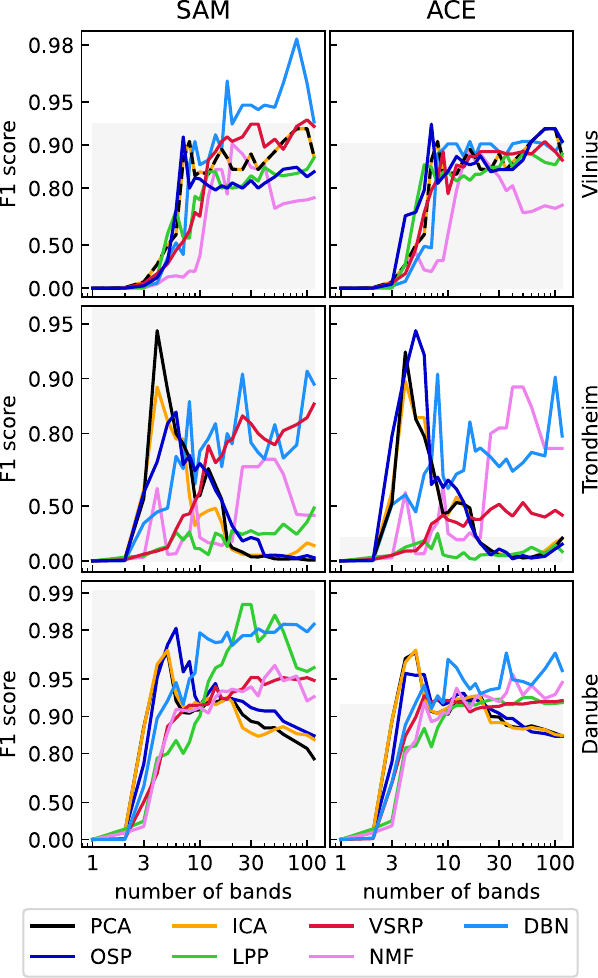}
    \caption{The F1 score of target detection using different numbers of bands. Only points above the shaded regions indicate parameters which exceeded the F1 score using all bands.}
    \label{fig:TD}
\end{figure*}

The parameters were standardized across the scenes. 
The spectrum used to characterize each target is the average spectrum of the target within the image. 
For ACE, the covariance matrix is generated from all the spectra of the image. 
Each component has its mean removed and its standard deviation is normalized to 1 before target detection is applied. 
The detectors are applied to the scenes for a varying number of bands.
In addition,   the SAM and ACE detectors are applied to the full set of bands on each scene. 
The F1 and intersection over union (IoU) scores were found to follow very similar patterns for all techniques, so only the F1 scores are plotted in Figure \ref{fig:TD}.
The threshold that maximized the F1 score was chosen for each technique. 

\subsection{Classification}

Classification is one of the most common applications of hyperspectral imaging. 
In classification, each pixel assigned to be a member of a class according to some criteria. 
The classification test uses a different dataset than the first five tests. 
This dataset was recently labelled in order to train on onboard classification algorithm, which, as of June 2023, is in daily use on the HYPSO-1 satellite~\cite{Roysland2023}. 
The dataset consists of 10 images from HYPSO-1 partitioned into 10 classes: forest, urban, rock, snow, sand, thick clouds, thin clouds, shadows, and 2 types of water. 
Any pixels which were saturated in any band were removed from the dataset before the processing.

\begin{figure*}[tb]
    \centering
    \includegraphics[width=0.6\textwidth]{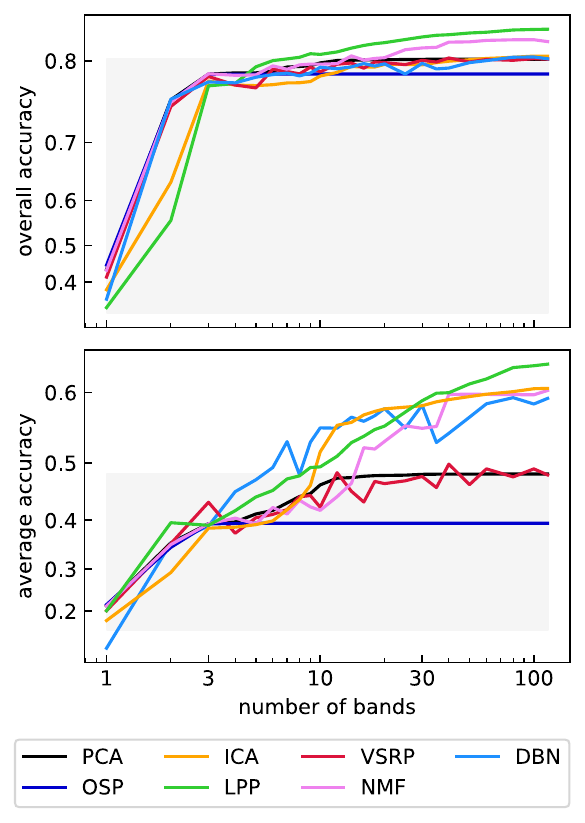}
    \caption{The overall accuracy (top) and average accuracy (bottom) of classification with a linear SVM classifier as a function of the number of bands selected, for the different DR techniques.}
    \label{fig:classification}
\end{figure*}

For the evaluation of the DR methods, a linear support vector machine (l-SVM) was employed. 
An l-SVM assigns a DR pixel, $x$, to one of two classes by performing a linear transformation and an offset
\begin{equation}
    C = \text{sign}(w^{T}x - b),
    \label{eq:SVM}
\end{equation}
where $w$ and $b$ are learned during the training process. 
For multiclass classification, several SVMs are joined together through a decision function. 
Here, they were trained using the one-versus-rest decision function, to maintain lower processing times than one-versus-one, at the cost of somewhat lower accuracy. 
For each method and each set of bands, the l-SVM was trained with $10^{4}$ pixels from the total set of about $2\times 10^{7}$ pixels. 
As in the target detection experiment, each component has its mean removed and has its standard deviation normalized to one. The classification is evaluated using two metrics, the overall and average accuracies (Fig. \ref{fig:classification}). 
Because of the variation in the size of the classes, there is considerable difference between these two metrics. 

%% file: Discussion.tex
\section{Discussion}
\label{sec:discussion}

The different DR algorithms behaved somewhat differently on each of the six tests. 
However, the timing experiments revealed one universal trend:  reducing the number of samples can vastly reduce execution time. 
While the execution time of LPP differed the most, by about 6 orders of magnitude between the smallest and largest sample sizes, even VSRP varied by almost 2 orders of magnitude, because of the time to compute the mean.  
Each of the techniques experienced a fairly consistent computation time for low sample sizes, during which computational overhead dominated the time, but around $10^{3}$ pixels, computational time began increasing. 
Most of the techniques scaled linearly with increasing sample size above that, but  LPP increased in proportion to $N^{2}$.

The computation time also varied with the number of computed bands, for those techniques for which the number of bands could be selected. 
OSP showed the most significant variation, almost 3 orders of magnitude variation for going from 1 to 116 bands. 
PCA with column sampling showed a reduced computation time relative to ordinary PCA when computing a reduced number of bands. 

The reconstruction accuracy likewise depends on the number of components used in the reconstruction. 
In general, the accuracy of all the methods improved when more components were used.
The best performance was achieved by PCA and FastICA, which is a rotation of PCA. 
OSP, NMF, and DBN achieved comparable reconstruction over much of the range, but reconstruction by NMF tends to stop improving above certain number of components. 
The components from LPP and VSPR reconstructed the scenes the worst of all. 
Much of the difference between those two methods is because the VSPR implementation removes the mean before separating the signal into components. 
The linear methods, by presenting and orthogonal basis for the vector space, are guaranteed to recreate the signals in the limit of infinitely many bands. 

Reducing the number of pixels incorporated into the calculation shows very little negative impact for the two most accurate reconstruction techniques, PCA and OSP. 
In fact, calculating with only 200 pixels gives approximately the same performance with OSP as calculating on the whole dataset. 
PCA shows some further improvement up until about $10^{3}$ pixels are used, but its performance  for less than $10$ components saturates even when only a fraction of the data is used. 
Notably, OSP exceeds the performance of PCA when only relatively few pixels are used to calculate the transformation, but many bands are included. 

The mutual information tests reveals three different groups of techniques. 
FastICA showed the least mutual information between components, and PCA showed only a little more. 
LPP, VSRP and DBN showed intermediate amounts of mutual information, while OSP and NMF components shared much mutual information. 
The positive result for FastICA is somewhat expected, since its objective function its designed to minimize mutual information.
For OSP, the first few components share less mutual information with the others than later components, and each component shares about the same amount of mutual information with all subsequent components. 

Imaging artifacts appeared most prominently in the components found by PCA and ICA.
ICA, in particular, finds many signals that are mostly artifact. 
On the other hand, the ATPV was consistently low for all OSP components, which was unexpected.
In addition, on the Vilnius and Danube scenes, OSP has the same level of ATPV as the median band from the whole image. 
As such, it seems that OSP can function as a benchmark for a normal amount of cross-track variance. 
The susceptibility of the other techniques seems to vary according to the scene. 
LPP, DBN, and NMF are more afflicted in the Vilnius scene than in the Trondheim scene.
VSRP is moderately affected in both scenes.
Because the ATPV is relatively high on the raw Danube scene, it is difficult to determine whether the large ATPV for all the DR techniques reflects artifacts or the structure of the scene itself. 

For target detection, we found that the SAM generally exceeded the accuracy of the ACE, contrary to previous findings. 
That could be due to the composition of the scenes, which were composed of land, water, and clouds (except Vilnius). 
Thus the covariance matrix was dominated by the contrast between those objects, so the subtler signals necessary to identify the particular water body were obscured. 
The performance of each technique varied significantly with the number of bands retained. 

No  DR technique proved optimal for target detection in all three scenes, but two groups which seemed to have similar behavior on each scene appeared. 
PCA, ICA, and OSP each achieved a performance peak at a low number of bands, but their performance decreased with an increasing number of bands, for each scene. 
On the Vilnius scene, these techniques achieve a second performance peak at over 50 bands, but their performance continues to deteriorate with an increasing number of bands on the other two scenes. 
The other hand, NMF, LPP, DBN, and VSRP generally show an increasing F1 score with number of bands, with some variation.


In the classification task, the LPP achieved the the highest results for both overall accuracy and average accuracy. 
It exceeded the other techniques on overall accuracy when 5 or more components were retained, although NMF performed nearly as well.
It exceeded them on average accuracy as well, when over 30 components were retained. 
DBN, NMF, LPP, and ICA exceeded the performance of classification using all bands without dimensionality reduction on both metrics. 
OSP reached a performance plateau at about 5 bands and gave the worst performance on both classification metrics. 

In summary, no one DR method proved preferable to the others on all tests, but some general patterns showing the suitability of each method to particular situations did emerge. 
PCA gave robust overall performance, with the lowest reconstruction error and good performance on the target detection and classification for low numbers of bands. 
The approximate column-sampling formulation is computed particularly fast. 
However, unlike LPP, DBN, NMF, and VSRP, it is not able to continue to extract useful information after a certain number of bands, here about 10. 
This is exemplified by target detection  and classification performance that either plateaus or decreases when more bands are included. 
Moreover, PCA and ICA are the most affected by the imaging artifacts. 

It is not unsurprising that ICA shares many characteristics with PCA, since its FastICA variant is a transformation of a reduced PCA subspace. 
The most notable differences from PCA appeared in the classification test, in which ICA outperformed PCA by about 10 percentage points in average accuracy when using over 10 bands. 
It also had the lowest mutual information between bands of any of the techniques. 
However, it takes longer to compute than PCA and is even more susceptible to the artifacts. 

OSP balances quick computation and good performance. 
Unexpectedly, OSP seemed to be nearly invulnerable to the imaging artifacts.
It also provided  fairly accurate reconstruction after ICA and PCA, and performs similarly to those two techniques on the supervised tasks. 
Its computational time requirement varies significantly depending on the number of bands computed. 
Together with NMF, it had the highest mutual information between bands, so in that regards it is somewhat different than PCA and ICA.

Although VSRP is less commonly applied to hyperspectral data than the aforementioned methods, it showed competitive performance. 
Of all the techniques, it was the fastest to compute for any number of bands and sample size. 
Moreover, it was only moderately affected by imaging artifacts and showed competitive performance on both supervised learning tasks. 

The computational time of LPP scaled extremely strongly with sample size, so that it required the most time of any technique for large sample sizes. 
However, its computational time was competitive for smaller sample sizes. 
It had the worst reconstruction error of all techniques, and had a moderate level of mutual information between bands. 
Nonetheless, it achieved the best performance on the classification task. 
LPP seems to include useful information in almost all bands, and so its performance does not saturate at low numbers of bands, unlike PCA, OSP, and ICA. 

The nonlinear techniques, NMF and DBN, show moderate overall performance. 
They both require relatively long computation time and have similar reconstruction error, despite the low reconstruction accuracy of NMF on the Vilnius scene. 
DBN shows somewhat less mutual information between bands than NMF and seems to be somewhat less susceptible to imaging artifacts. 
DBN shows better performance than NMF on the target detection task, while NMF performs better on classification.

All together, the experiments show that calculating the DR transformations from a small number of pixels randomly chosen from the scene can notably accelerate the computation of DR components, and reduces the reconstruction accuracy only a few percent. 
Calculating with only some of the pixels thus could be critical to meeting on-board processing constraints. 
There is not a overall correlation between either the reconstruction error or the mutual information of the components and the performance on the supervised tasks.
For example, LPP shows the worst reconstruction error but the highest OA and AA in classification. 
However, it does seem like those methods which have a relatively low reconstruction error for few bands (ICA, OSP, PCA), perform well at the supervised tasks if only a few bands are considered, but their performance plateaus or decreases when more bands are included. 
Since they bring so much of the variance into the first few components, there is little signal left to distribute as more bands are included. 


%% file: Conclusion.tex
\section{Conclusions}
\label{sec:conclusion}

The importance of onboard processing is growing as more platforms are developed for remote hyperspectral imaging. 
The need is particularly great as the platforms become more autonomous and the size of their data downlink rate relative to the amount of data they collect decreases. 
The experiments presented here draw out some distinguishing characteristics of different types of DR algorithms which are enumerated below.

\begin{enumerate}
    \item There is no dimensionality reduction algorithm which outperforms the others on all tests.
    \item All the transformations can be calculated more quickly and reasonably well using only a small subset of the pixels. For example, the maximum reconstruction accuracy seemed to be achieved when using only about 200 pixels with PCA and OSP. The time required by most of the methods seems to increase linearly with the number of pixels included. The time required by LPP, on the other hand, seems to increase quadratically.  
    \item OSP and NMF compute components which share the most mutual information, which seems to indicate a redundancy between the components. However, this does not seem to have any systematic detrimental effect on the reconstruction error.
    \item A low amount of mutual information between components seems to predict susceptibility to imaging artifacts. The two methods with the least mutual information between components, PCA and ICA, are also the most susceptible to imaging artifacts. Of those with the most MI between components, OSP seems to be uniquely robust against the striping artifact, while NMF is not.
    \item For OSP, PCA, and ICA, target detection performance peaks with a small number of bands and then decreases. The other techniques seem to require more bands to achieve similar performance, but their performances do not decrease as more bands are incorporated.
    \item The target detection accuracy depends as much on the number of bands selected as it does on the choice of the method. It also depends on the choice of the scene.
    \item For three or fewer bands, the different methods perform similarly in classification. As more bands are included, however, LPP exceeds the other methods. NMF, DBN, and ICA also exceed the performance of the raw data, PCA and VSRP match it, and OSP performs worse. 
\end{enumerate}

The numerical experiments also bring forth a number of new questions which could be studied in more detail. 

\begin{enumerate}
    \item The relationship between artifact susceptibility and mutual information should be investigated in more detail. If the relationship is sufficiently general, mutual information could be used as a metric to characterize or search for artifacts. 
    \item A more thorough theoretical analysis of the interactions between algorithms and scene statistics could help to better understand the effect of DR on target detection and classification. It could aim to predict when SAM outperforms ACE, contrary to expectations, e.g. \cite{manolakis2013detection}.
    \item Methods for accelerating the algorithms could be investigated in more detail. Notably, column sampling improves the speed of PCA by about an order of magnitude, and there likely exist similar ways of accelerating the other techniques. The non-linear methods, which excel on metrics other than speed, might also be accelerated by finding more appropriate initialization conditions. 
    \item Most of the analysis presented here relies on single images, except for the classification test. Future analyses could investigate how different DR techniques can be incorporated into pipelines for continuously-observing platforms.
\end{enumerate}